# Beyond Gamification: Implications of Purposeful Games for the Information Systems Discipline

*Kafui Monu and Paul Ralph*

*Working Paper*

## Introduction

In late 2011, players of a multiplayer online puzzle game, Foldit, were delighted to discover that they had not only won the game, but that their epic performance had solved "the crystal structure of M-PMV retroviral protease" by generating "models of sufficient quality for successful molecular replacement and subsequent structure determination" providing "new insights for the design of antiretroviral drugs" (Khatib et al. 2011, p. 1).

Foldit was developed by game designers and biochemists to recruit the help of gamers to understand protein folding in DNA structures. Foldit uses 3D graphics to depict DNA puzzles and gamers worldwide compete in solving the puzzle (Khatib et al. 2011). The solutions were studied by biochemists for the development of new enzymes (Bourzac 2008). Foldit's developers leveraged many non-scientists' passion for puzzles to make genuine breakthroughs, which had previously eluded scientists. One biochemist explained "I worked for two years to make these enzymes better and I couldn't do it. Foldit players were able to make a large jump in structural space and I still don't fully understand how they did it" (Marshall and Nature Magazine, 2012).

Foldit is one of many examples of using games in non-game contexts. The success of such games comes alongside the introduction of *gamification*, the process of applying game design elements such as points, badges, competition, positive feedback and interactivity to non-game contexts including marketing, education and research (Deterding et al. 2011). For instance, Duolingo(.com) helps users learn a second language using progressively more difficult translation exercises, feedback mechanisms, challenges and trophies. Progress is visualized using a "skill tree" and trophies may be shared on Facebook to increase the sense of social accomplishment. Fundamentally, gamification, and the use of games in non-game contexts in general, is based on the presentation, manipulation, storage and analysis of information and is becoming an important design principle for Information Systems (IS) (Schell, 2010).

Gamification is becoming an important way of engaging information system users and IS researchers should study the area (Kankanhalli et al. 2012), but there are other ways in which games are used in non-game contexts. For instance, full-fledged "serious" games, similar to Foldit, and simulations have a long history of being used to train and educate managers (Keys and Wolf 1990). We refer to serious games, simulations, gamification, and any other form of play for non-entertainment reasons, as *purposeful gaming*.

Since there are several types of purposeful games, it is important to know whether there are situations where one type of game is better than another and to identify all possible types of games, beyond gamification, serious games, and simulation. But most importantly, since most examples of purposeful gaming uses some form of information technology (Schell 2010) it is important to know how IS literature is changed, validated, and expanded by purposeful gaming, beyond just gamification.

This paper therefore explores the following question: *What are the implications of purposeful games for the IS discipline?*

We discuss the current state of purposeful gaming (§2). Section three shows how purposeful gaming affects the core of the IS discipline. Section four provides the dimensions of purposeful gaming that are relevant to the IS Discipline. Lastly, section five summarizes the paper's contribution and discusses possible avenues for IS/purposeful gaming research.

## 2. The Current State of Purposeful Gaming

"Gamification" describes the relatively new and popular form of purposeful games, having only emerged in the popular lexicon in September 2010 (cf. http://www.google.com/insights/search/#q=gamification) after its introduction by the digital media industry (Deterding et al. 2011). It has grown quickly into a billion dollar industry (Gartner Group 2012) and has been specifically important in the new area of mobile computing (Zichermann and Cunningham 2011). However, it draws on other purposeful games; consequently, this section reviews three key themes related to purposeful games – *serious games*, gamification itself, and *gameful design*.

"Serious games" are "games used for purposes other than mere entertainment" (Susi et al. 2007, p. 1). Gamification is distinct from serious games since applying game elements does not necessitate creating a game. For instance, a company gamifying its sales experience by providing points to customers for purchases is not creating a serious game as it does not include enough game elements for most people to classify it as a game. Serious games have been used in a variety of educational contexts including military, law enforcement, ethics, firefighting, aviation, health and safety and mental health (Michael and Chen, 2005). Serious games also include games aimed at scientific discovery (e.g., Foldit), increasing employee productivity, or selling products.

Indeed, most current gamification efforts have been centered on marketing efforts and increasing brand awareness (Zichermann & Cunningham, 2011). In this context, gamification is a translation of business objectives into desired customer or employee behaviors, a points system attached to desired behaviors, and enhancement of the experience with badges ("achievements"), leaderboards, challenges, quests, social elements and rewards. This pattern led Schell (2010) to predict a "Gamepocalypse"– a dystopic future where everyone earns meaningless points for every activity from watching television to shopping to riding the bus. Schell worries that gamification will empower amoral advertisers to manipulate the masses into whatever behaviors are most profitable.

In contrast, McGonigal (2011) hypothesizes that multiplayer adventure games, including World of Warcraft, simulate the ideal environment for human performance – a player embarks on an epic quest, has complete control over her actions and is provided with everything necessary to become proficient in the required skills. These characteristics correspond exactly to the primary antecedents of motivation – purpose, autonomy and mastery (Pink 2009). McGonigal further argues that redesigning real-world social structures to mimic those of adventure games will improve our collective productivity and capacity to solve global problems.

Somewhere between these extreme views, Deterding et al. (2011) suggests that gamification illuminates the idea of "gamefulness". While *play* connotes open-ended, improvised and free interaction, *game* connotes goal-oriented activity within a systems of rules; consequently, one can design for *playfulness* (e.g., classic toys including balls, dolls and skipping ropes) or *gamefulness* (e.g., classic board games including Scrabble, Risk and Battleship). Considering gamefulness as a design lens changes the framing of our discussion from gamification as a marketing technique to gamefulness as a mechanism for improving information system development (ISD). ISD could use the ideas of gamefulness to design information systems that "act" like games. The results of using gameful design are purposeful games.

## 3. Purposeful Games are an IS Topic

Purposeful gaming is an important topic for IS research for three reasons – many traditional IS are being gamified, gamifying organizational work entails substantial ISD, and the use of games in business has important implications for core IS theories. We elaborate each of these below.

### *3.1 Gamified Management Information Systems*

As mentioned, many types of management information systems are being gamified. This section discusses the implications for IS research of gamifying two types of information systems central to the IS discipline (Baskerville & Myers, 2002) – customer support systems (customer-facing) and training systems (employee-facing).

The purpose of a customer support or customer relationship management system (CRM) is to support relationship marketing, i.e., creating a personal connection between the company and its customers through value creation, multi-channel integration, and information management (Payne & Frow, 2006). Traditionally, CRMs are used only in the information management process and as part of the multi-channel integration process. However, a gamified CRM may also create value. For example, Nike's Nike+ app not only helps Nike gather detailed information about runners' habits but also helps runners track their performance. Nike+ used game design elements including progression and competition simultaneously to attract users, create value and collect and organize additional consumer data (Zichermann & Cunningham, 2011). In this way, gamifying the CRM gave it direct value for customers, rather than simply enabling information management and value delivery.

Similarly, the purpose of (automated) training systems is to facilitate skill improvements among users. Training systems may reduce cost, improve tracking, convenience and training consistency and provide not only information but also "practice, feedback, and guidance" (Welsh et al., p. 249). However, optional courses have low completion rates among employees (Welsh et al., 2003). Gamifying training systems may increase completion rates by improving employee motivation (Ryan et al. 2006). As mentioned previously, training systems have long been a focus of serious games. For example, ERPSim teaches students to use SAP to make strategic decisions in a co-operative game environment (Cronan et al., 2011).

### 3.2 The Crucial Role of Information System Development in Gamification

A basic method for gamifying organizational work might consist of defining business objectives; deriving desired user behaviours; devising extrinsic rewards that appeal to users; and implementing game mechanics including points and trophies to connect desired behaviours to extrinsic rewards (Zichermann and Cunningham 2011). Some researchers have argued that this method is not sustainable because it only focuses on extrinsic rewards (Nicholson 2013). However, the basic method illustrates the relationship between gamification and ISD – the system that connects desired behaviours to rewards is clearly an information system. Therefore, gamifying organizational work entails ISD.

Put another way, purposeful games are a type of ISD project and have unique ISD concerns, since iterative prototyping, "play-testing" and trial and error (Stacey and Nandhakumar 2008) are how games are designed (Schell 2008). Defining specific requirements for purposeful games is difficult, or impossible, due to our limited ability to predict how fun and engaging purposeful games will be (Schell 2008).

More generally, ISD involves understanding the domain, and designing the system based on this domain understanding (Jarke et al. 1993). An added concern for the purposeful game designer is that technologies support both the work process and the game (Zichermann and Cunningham 2011). To the best of our knowledge, little or no research has addressed how game elements should be considered during information system development and how work and game elements interact with each other in ISD.

Another issue with gamified ISD is that the user is extremely important, since the experience is the result of the interaction between the game and the player (Schell 2008). So when designing gamified IS it may be important to focus less on concepts like, objects and events and more on agents and actors. And developers may need to employ system analysis and design techniques that focus on these concepts (Monu 2011).

### 3.3 IS Theories of Gamified Information System

Purposeful games also have implications for core IS theories. For example, current IS use theories may not be appropriate for gamified information systems. The theory of task-technology fit (TTF) posits that task and technology characteristics interact to create task-technology fit, which affects use and performance (Goodhue and Thompson 1995). But this fit theory is not appropriate for purposeful games because game characteristics, which affect whether people engage in a game (Rouse and Ogden 2005), are different from both task and technology characteristics.

Similarly, the Unified Theory of Acceptance and Use of Technology (UTAUT) posit that IS adoption and use are governed by performance expectancy, effort expectancy, social influence and behavioural intention (Venkatesh et al. 2008). In contrast, adoption and use of video games are governed by factors including autonomy (the ability to use the tools in the game to do what the player wants), presence (immersion in the game world) and perceived competence (opportunities for positive feedback and acquiring skills) (Ryan et al. 2006). While perceived competence is related to performance expectancy and autonomy is related to effort expectancy, presence is not represented in the UTAUT. Moreover, although Venkatesh et al. (2008) showed that attitude toward the technology does not affect behavioural intention when performance and effort expectancy are considered, attitudinal constructs may more powerfully affect use intention in gamified IS as positive attitude created during play is a core purpose of games (Caillois 2005).

For instance, if the same task and technology but different game mechanics were used would Foldit be as effective? What if Foldit only had specific challenges and quests? Or had a skill tree mechanic? Do these mechanics fit the non-game context of protein manipulation better than leaderboards? Game-task-technology fit could affect the utilization and performance impact of Foldit but without separating the game characteristics from the task characteristics it is impossible to test how the game mechanics affect the utilization and performance of the gamified IS.

## 4. Conceptual Dimensions of Purposeful Games

As mentioned, there are several types of purposeful games such as, serious games, gamification, and simulations. Some researchers have even made distinctions between types of gamification that use different game elements (Nicholson 2013). Since there are can be many types of purposeful games, and these types may interact with IS differently, we propose a set of conceptual dimensions that can classify gamified information systems for IS research.

To determine these dimensions, we can think of a specific information system as a point in a multidimensional design space. When an information system is modified, its position in the design space is altered. One way to conceptualize purposeful games is as information systems that have characteristics that exist in this design space, based on particular dimensions. We see at least three dimensions of purposeful games that are relevant to the IS discipline (See Figure 1).

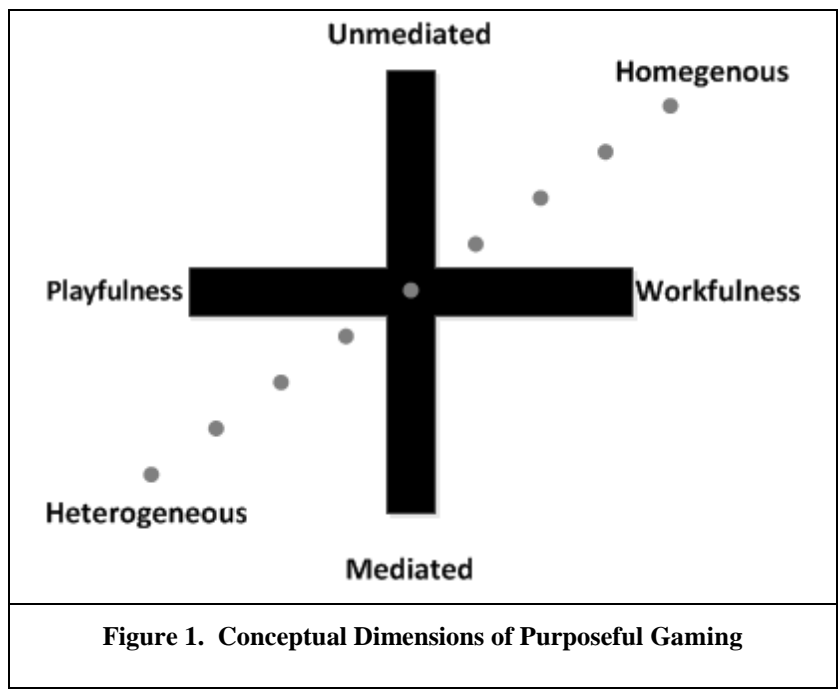

**Figure 1. Conceptual Dimensions of Purposeful Gaming**

The first dimension is playful vs. workful. Deterding et al. (2011) makes a distinction between playfulness and gamefulness by stating that whereas playing "denotes a more free-form, expressive, improvisational, even "tumultuous recombination of behaviors and meanings", gaming "captures playing structured by rules and competitive strife toward goals" (p. 11). We chose playful rather than gamefulness as a dimension because Caillois (2005) defines a game as a rule-bound play activity with specific goals.

Caillois (2005) defines play as a non-obligatory activity, set in a pre-determined place and time; separate from reality, which has uncertain consequences and is unproductive. The only difference between play and games is that games are bound by rules while "pure play" is "make-believe" (Caillois 2005). Work on the other hand is defined as "human activities, [usually] within the context of formal organizations, performed with the intention of producing something of acknowledged social value." (Rice et al. 1985, pp. 296). Kahn (1981) distinguishes play from work by the intention of whether the person wants to create something of social value through the activity (work) or perform the activity for its own sake (play).

This means that gamifying organisational work involves traversing the design space from workful to playful by giving the system more properties associated with games. So for instance, many management training games try to recreate the activities that exist in work but make sure that nothing of value is created and that the participants are engaging in it for experience (Keys and Wolf 1990). One can also add elements of work to games to "workify" game experiences. For instance, play becomes work for professional sports players (Caillois 2005) and, in the case of information technology the "gold-farming" industry turns video game players into workers who receive wages for entering game worlds (Dibell 2007).

The second dimension is between unmediated vs mediated. Games and work vary on their degree of being mediated through an information technology. Team sports and board games are predominately physical, while video games are mediated through information technology. Similarly work may be predominately physical (e.g., stocking shelves at a supermarket) or mediated through business information systems (e.g., trading derivative investments through electronic exchanges). Current gamification usually includes adding virtual elements (badges, leaderboards, etc.) to predominately physical work systems (Zichermann and Cunningham 2011) but it can also involve adding physical elements to predominately mediated work. For instance, Microsoft software developers play a physical card game called "Elevation of Privilege" that encourages and teaches them to model security threats to their code (Shostak).

The third dimension is homogeneous vs. heterogeneous. In games, people have diverse motivations including collecting, competing, exploring, achieving, directing, storytelling and humor (Klug and Schell 2006; Schell 2008). Consequently, games vary tremendously in the play motivations to which they appeal. For instance, real-time strategy games do not motivate storytelling as well as role-playing games (Rouse and Ogden 2005). Successfully gamifying work for diverse individuals may require appealing to diverse player motivations. For example, DuoLingo(.com) motivates achieving (leaderboards), exploration (discovery of new exercise and skills), and collecting (gaining points and badges), while Foursquare(.com) only motivates collecting through its badges. There are also different elements of organisational work; which include customers, participants, work practices, technologies, information and product and services (Alter 2006). Purposeful games may affect one type of work element or several. For instance, training games such as ERPSim focus on training employees or participants of the work system (Cronan et al. 2011), while Foldit directly relates to work practices, manipulating models of protein structure, and the product of the work practice, the structure of new enzymes (Khatib et al. 2011).

Each of these dimensions interacts with the artifacts, design, and theories of IS. The implication of the first dimension, workful vs. playful, is discussed in Section 2. In short, the artifacts, designs, and theories discussed in IS may not be appropriate for purposeful games and game research will need to be incorporated into the IS discipline to better understand purposeful games. The implication of the second dimension, unmediated vs. mediated, is that work in the IS discipline can be used to provide a better understanding of mediated purposeful games. And lastly, the implication of the homogenous vs. heterogeneous dimension is that more research is needed to study the interaction between game and work elements. The number of different work and game elements may affect the quality of the gamified information system. For instance, it may be that gamification that has heterogeneous work elements and homogenous game elements is as effective as gamification that has heterogeneous game elements. To gain a better understanding of purposeful games the IS discipline may not only have to incorporate game research, but also determine how it interacts with current research in business information systems.

# 5. Discussion and Conclusion

As purposeful games are rarely studied and under-theorized, the area is ripe for groundbreaking empirical research which should be conducted by researchers in the IS discipline. At least three broad research streams are evident: definitions, effectiveness, and design.

First, many of the terms surrounding purposeful gaming are not clearly defined or demarcated. For example, what is the relationship between gamification and serious games? Does the former create the latter if taken far enough? The conceptual dimensions in Section 4 provide the framework to discuss different purposeful games but more research is needed to understand how the dimensions can be used to define all forms of purposeful games.

Second, although various game design elements are recommended for non-game contexts, the relationships between each element and overall effectiveness is unclear. For example, which game design mechanic is more important for motivation – trophies or progression? A holistic understanding of gamification phenomena will therefore require one or more variance theories which identify the antecedents of gamification effectiveness, where gamification effectiveness may include different dimensions in different contexts; e.g., adoption and use in CRM; student engagement in education. This theory would provide constructs about effective gamification and a basis for more varied gamification practices. The theory would also contributes to IS research by providing theoretical constructs for gamified IS use, constructs that directly relate purposeful games to IS, and provide the basis for metrics of effective use of IS in purposeful gaming.

Third, gamification combines aspects of game design, software design, and ISD. The academic discourse includes starkly different descriptions of these processes; e.g., game design is dominated by iterative play-testing (Stacey and Nandhakumar 2008) and software design is dominated by the coevolution of problem frames and solution concepts (Ralph 2010) while ISD is often thought of in terms of lifecycle models (Ralph 2011). Understanding purposeful gaming will therefore require one or more process theories that explain how participants gamify existing systems in practice.

The dimensions should also act as important experimental variables for purposeful game research in IS, specifically business information systems. For instance, empirical work could be conducted on the use of avatars to create social presence in gamified information systems that vary in their degree of playfulness.

The contribution of this work is the discussion of the implications of purposeful gaming on the IS discipline. We identified work beyond gamification, such as serious games and simulations to gain a better understanding of how games will affect our field. The conceptual dimensions provide IS researchers and practitioners with a tool to classify and understand the wide spectrum of purposeful games.